\RequirePackage{fix-cm}
\documentclass[smallextended]{svjour3}       % onecolumn (second format)
\smartqed  % flush right qed marks, e.g. at end of proof
\usepackage{graphicx}
\PassOptionsToPackage{hyphens}{url}%\usepackage[hidelinks]{hyperref}
\usepackage{amsmath}

\usepackage{pifont}
\newcommand{\cmark}{\ding{51}}
\newcommand{\xmark}{\ding{55}}

\newcommand\ceil[1]{\lceil#1\rceil}

\usepackage[binary-units=true]{siunitx}

\usepackage{adjustbox}
\usepackage{booktabs,multirow}

\usepackage[inline]{enumitem}
\newlist{mylist}{enumerate*}{1}
\setlist[mylist]{label=(\roman*)}

\usepackage{array}
\newcolumntype{M}[1]{>{\centering\arraybackslash}m{#1}}
\usepackage[dvipsnames]{xcolor}
\definecolor{mygray}{gray}{0.4}
\newcommand{\light}[1]{{\fontsize{6}{6}\selectfont\textcolor{mygray}{#1}}}

\usepackage[pscoord]{eso-pic}
\newcommand{\placetextbox}[3]{% \placetextbox{<horizontal pos>}{<vertical pos>}{<stuff>}
  \setbox0=\hbox{#3}% Put <stuff> in a box
  \AddToShipoutPictureFG{% Add <stuff> to current page foreground
    \put(\LenToUnit{#1\paperwidth},\LenToUnit{#2\paperheight}){\vtop{{\null}\makebox[0pt][c]{\textcolor{red}{#3}}}}%
  }%
}%

\begin{document}

\placetextbox{0.5}{0.99}{\large\texttt{\parbox{1.2\textwidth}{This is a pre-print of an article published in Multimedia Tools and Applications. The final authenticated version is available online at: \href{https://rdcu.be/b8gzW}{https://doi.org/10.1007/s11042-020-09905-3}}}}%

\title{\MakeLowercase{gpu}RIR: A Python Library for Room Impulse Response Simulation with GPU Acceleration
\thanks{This work was supported in part by the Regional Government of Aragon (Spain) with a grant for postgraduate research contracts (2017-2021) co-funded by the Operative Program FSE Aragon 2014-2020. \newline This material is based upon work supported by Google Cloud.}%Grants or other notes about the article that should go on the front page should be
}

\titlerunning{\MakeLowercase{gpu}RIR: A Python Library for RIR Simulation with GPU Acceleration}        % if too long for running head

\author{David Diaz-Guerra       \and
        Antonio Miguel          \and
        Jose R. Beltran         %etc.
}


\institute{David Diaz-Guerra, Antonio Miguel and  Jose R. Beltran \at
              Department of Electronic Engineering and Communications \\
              University of Zaragoza, Spain %  \\
           \and
           David Diaz-Guerra \at
              \email{ddga@unizar.es} 
}

\date{Received: date / Accepted: date}

\maketitle

\begin{abstract}
The Image Source Method (ISM) is one of the most employed techniques to calculate acoustic Room Impulse Responses (RIRs), however, its computational complexity grows fast with the reverberation time of the room and its computation time can be prohibitive for some applications where a huge number of RIRs are needed. In this paper, we present a new implementation that dramatically improves the computation speed of the ISM by using Graphic Processing Units (GPUs) to parallelize both the simulation of multiple RIRs and the computation of the images inside each RIR. Additional speedups were achieved by exploiting the mixed precision capabilities of the newer GPUs and by using lookup tables. We provide a Python library under GNU license that can be easily used without any knowledge about GPU programming and we show that it is about 100 times faster than other state of the art CPU libraries. It may become a powerful tool for many applications that need to perform a large number of acoustic simulations, such as training machine learning systems for audio signal processing, or for real-time room acoustics simulations for immersive multimedia systems, such as augmented or virtual reality.
\end{abstract}

\section{Introduction}
\label{sec:intro}

The simulation of the acoustics of a room is needed in many fields and applications of audio engineering and acoustic signal processing, such as training robust Speech Recognition systems \cite{weng_recurrent_2014} or training and evaluating Sound Source Localization \cite{griffin_localizing_2015} or Speech Enhancement \cite{williamson_time_2017} algorithms. Although there are many low complexity techniques to simulate the reverberation effect of a room in real time, as the classic Schroeder Reverberator \cite{schroeder_natural_1962}, some applications require an accurate simulation of the reflections causing the reverberation. The information of all those reflections is gathered in the Room Impulse Response (RIR) between the source and the receiver positions, which allows to simulate the reverberation process by filtering the source signal with it. Our goal in this work is to provide a fast method to obtain these RIRs.

The Image Source Method (ISM) is probably the most used technique for RIR simulation due its conceptual simplicity and its flexibility to modify parameters such as the room size, the absorption coefficients of the walls, and the source and receiver positions. We can simulate any level of reverberation by modifying the room size and the absorption coefficients, but the computational complexity of the algorithm grows fast as the number of reflections to simulate increases. In addition, many applications require the computation of multiple RIRs for several source and receiver positions, e.g. to simulate a moving source recorded with a microphone array. Furthermore, with the increasing popularity of Machine Learning techniques, the need for computing randomly generated RIRs on the fly for huge datasets in a reasonable time is constantly increasing.

%Firstly developed to support the graphics computations of video-games, Graphics Processing Units (GPUs) are today one of the best and cheapest ways to increase the speed of many algorithms that can be expressed in a parallel form. Despite parallelizing most of the stages of the ISM is quite straightforward, to the best of our knowledge, only \cite{fu_gpu-based_2016} propose to implement it in GPUs. In this paper we  present a new GPU implementation with a higher degree of parallelization, better performance, and, most importantly, we provide it as a free and open-source Python library to be used by the research community\footnote{The code, the documentation, the installation instructions, and examples can be found in \url{https://github.com/DavidDiazGuerra/gpuRIR}}. Using our library does not require any knowledge about GPU programming, but just having a CUDA compatible GPU and the CUDA Toolkit and to install the library and use it as any RIR simulation library.

Firstly developed to support the graphics computations of video-games, Graphics Processing Units (GPUs) are today one of the best and cheapest ways to increase the speed of many algorithms that can be expressed in a parallel form. Despite parallelizing most of the stages of the ISM is quite straightforward, to the best of our knowledge, only \cite{fu_gpu-based_2016} proposed to implement it in GPUs. Although they showed that using GPUs it was possible to speed-up the RIR simulations, they did not provide the code of their implementation and the acoustic signal processing and audio engineering communities have not embraced their approach. In addition, they used an overlap-add strategy with atomic operations to combine the contributions of each image source, which strongly reduces the level of parallelism. In this paper, we  present a new GPU implementation with a higher degree of parallelism, which allows us to achieve higher speed-ups with cheaper GPUs. Motivated by the performance boost obtained with the use of lookup tables (LUTs) in the CPU implementations, we also study its use in our GPU implementation. Finally, we propose a 16-bit precision implementation which can increase even more the simulation speed in the newer GPUs with mixed precision support. 

Table \ref{tab:libraries} shows some state of the art implementations of the ISM and compare some of their main characteristics. We can see how our implementation is the only one with GPU acceleration that is available as a free and open source library\footnote{The code, the documentation, the installation instructions, and examples can be found in \url{https://github.com/DavidDiazGuerra/gpuRIR}} and how it includes some features (further explained in section \ref{sec:improvements}) that are not included in other Python libraries. Using our library does not require any knowledge about GPU programming, but just having a CUDA compatible GPU and the CUDA Toolkit, and it can be installed and used as any CPU RIR simulation library. 

\begin{table}%[tbp]
\caption{Comparison of some state of the art ISM implementation}
\label{tab:libraries}
%\begin{adjustbox}{width=1\textwidth}
\begin{tabular}{llllll}
    \hline\noalign{\smallskip}
    & RIR generator \cite{habets_room_2010}     & pyroomacoustics \cite{scheibler_pyroomacoustics_2018}   & \cite{lehmann_diffuse_2010}     & \cite{fu_gpu-based_2016}  & gpuRIR\\
    \noalign{\smallskip}\hline\noalign{\smallskip}
    Open source library (language)  & \cmark (Matlab and Python)   & \cmark (Python)    & \cmark (Matlab)   & \xmark  & \cmark (Python) \\
    Implementation language             & C++       & Python and C++    & Matlab    & CUDA      & CUDA   \\
    Fractional delays                   & \cmark    & \cmark            & \cmark    & \cmark    & \cmark \\ 
    Negative reflection coefficients    & \xmark    & \xmark            & \cmark    & \xmark    & \cmark \\ 
    Diffuse reverberation model         & \xmark    & \xmark            & \cmark    & \xmark    & \cmark \\ 
    GPU acceleration                    & \xmark    & \xmark            & \xmark    & \cmark    & \cmark \\ 
    Lookup table implementation         & \xmark    & \cmark            & \xmark    & \xmark    & \cmark \\
    Mixed precision implementation      & \xmark    & \xmark            & \xmark    & \xmark    & \cmark \\
    \noalign{\smallskip}\hline
\end{tabular}
%\end{adjustbox}
\end{table}

The contributions of the paper are the following:
\begin{mylist}
    \item we present a new parallel implementation of the ISM which fits better with the newer GPUs architectures than the only alternative available in the literature,
    \item we discuss how to increase the performance of GPU programs with several techniques such as using Lookup Tables or 16-bit precision floating point arithmetics,
    \item we present a new Free and Open Source Python library exploiting this implementation, and
    \item we compare it against several state of the art ISM implementations and show how ours is two orders of magnitude faster than them.
\end{mylist}

The reminder of this paper is structured as follows. We review the ISM in section \ref{sec:ISM}, section \ref{sec:implementation} explains how we have parallelized it, and section \ref{sec:library} presents the Python library. Finally, in section \ref{sec:results}, we compare the performance of our library against three of the most commonly used RIR simulation libraries and section \ref{sec:conclusions} concludes the paper.

\section{The Image Source Method (ISM)}
\label{sec:ISM}

The Method of Images has been widely used in many fields of physics to solve differential equations with boundary conditions, but its application for RIR estimations was originally proposed by Allen and Berkley \cite{allen_image_1979}. In this section, we first review their original algorithm and then explain some of the improvements that have been proposed to improve both its accuracy and computational performance.

\subsection{Original Allen and Berkley algorithm}

\begin{figure}[]
    \centering
    \includegraphics[width=0.75\textwidth]{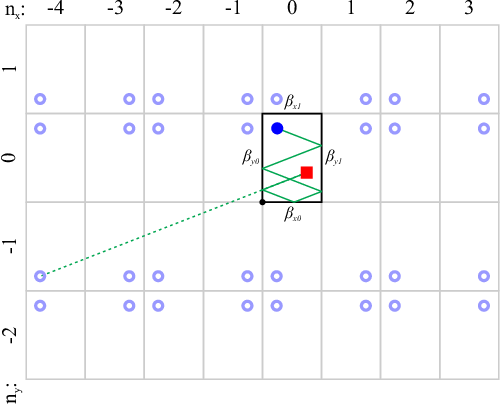}
    \caption{Image sources for a two dimensional room. The red square and the blue dot represents the receiver and the source and the blue circumferences represents the image sources. The solid green line represents one of the multiple reflection paths and the dashed green line the direct path of the equivalent image source. The black dot is the origin of the coordinates system.}
    \label{fig:ISM}
\end{figure}

The main idea behind the ISM is to compute each wave-front that arrives to the receiver from each reflection off the walls as the direct path received from an equivalent (or image) source. In order to get the positions of these image sources, we need to create a 3D grid of mirrored rooms with the reflections of the room in each dimension; as shown in Fig. \ref{fig:ISM} simplified to 2D for an example. 

If the number of images we want to compute for each dimension are $N_x$, $N_y$ and $N_z$, then we define a grid $\mathcal{N}$ of image sources $\textbf{n} = (n_x, n_y, n_z) :$ $\ceil{-N_x/2} \leq n_x < \ceil{N_x/2}, \ceil{-N_y/2} \leq n_y < \ceil{N_y/2}$ and $\ceil{-N_z/2} \leq n_z < \ceil{N_z/2}$ (where $\ceil{\cdot}$ stands for the round toward positive infinity operator). The coordinates of the position of each image $\textbf{p}_\textbf{n}=(x_\textbf{n}, y_\textbf{n}, z_\textbf{n})$ are calculated using its grid indices, the position of the source and the dimensions of the room; as an example, the component x would be calculated as
\begin{equation}
    x_\textbf{n} = \begin{cases}
        n_x L_x + x_s & \text{if $n_x$ is even} \\
        (n_x+1) L_x - x_s & \text{if $n_x$ is odd}
    \end{cases},
\end{equation}
where $\textbf{L}=(L_x, L_y, L_z)$ is the size of the room and $\textbf{p}_s=(x_s, y_s, z_s)$ is the position of the original source. The $y$ and the $z$ coordinates can be obtained similartly.

The distance $d_\textbf{n}$ from the image source $\textbf{n}$ to a receiver in the position $\textbf{p}_r=(x_r, y_r, z_r)$, and therefore the delay of arrival $\tau_\textbf{n}$, is trivial if we know the image source position:
\begin{equation}
    d_\textbf{n} = ||\textbf{p}_r-\textbf{p}_s||,
\end{equation}
\begin{equation}
    \tau_\textbf{n} = \frac{d_{\textbf{n}}} {c},
    \label{eq:tau}
\end{equation}
where $||\cdot||$ denotes the Euclidean norm and $c$ is the speed of sound.

In order to calculate the amplitude with which the signals from each image source arrive to the receiver, we need to take into account the reflection coefficients of the walls of the room. We define $\beta_{x0}$ as the reflection coefficient of the wall parallel to the $x$ axis closest to the origin of the coordinates system and $\beta_{x1}$ as the farthest; $\beta_{y0}$, $\beta_{y1}$, $\beta_{z0}$ and $\beta_{z1}$ are defined equivalently. Finally, if we define $\beta_\textbf{n}$ as the product of the reflection coefficients of each wall crossed by the path from the image source $\textbf{n}$ to the receiver, its amplitude factor will be
\begin{equation}
    A_\textbf{n} = \frac {\beta_{\textbf{n}}} {4\pi \cdot d_{\textbf{n}}}.
    \label{eq:amp}
\end{equation}

Knowing the amplitude and the delay for each image, we can easily obtain the RIR as the sum of the contribution of each image source:
\begin{equation}
    \label{eq:ISM}
    h(t) = \sum_{\textbf{n} \in \mathcal{N}} A_\textbf{n} \cdot \delta(t - \tau_\textbf{n}),
\end{equation}
where $\delta(t)$ is the Dirac impulse function.

\subsection{Improvements to the original algorithm}
\label{sec:improvements}

\subsubsection{Fractional delays}
In order to implement \eqref{eq:ISM} in the digital domain, we need to deal with the fact that the values of $\tau_\textbf{n}$ may not be multiples of the sampling period. The original algorithm proposed to just approximate the fractional delays by the closest sample, however, the error introduced by this approximation is too high for some applications, such as Sound Source Localization with microphone arrays. In \cite{peterson_simulating_1986}, Paterson proposed to substitute the Dirac impulse function by a sinc windowed by a Hanning function:
\begin{equation}
    \label{eq:sinc}
    \delta'(t) = \begin{cases}
        \frac{1}{2} \left(1+\cos{\frac{2\pi t}{T_\omega}}\right) \text{sinc}(2\pi f_c t) & \text{if} -\frac{T_\omega}{2}<t<\frac{T_\omega}{2} \\
        0 & \text{otherwise}
    \end{cases},
\end{equation}
where $f_c$ is the cut-off frequency, $T_\omega$ is the window length, and the sinc function is defined as $\text{sinc}(x)=\sin(x)/x$. This is motivated by the low pass anti-aliasing filter that would be used if the RIR was recorded with a microphone in the real room. A window duration of $T_\omega=\SI{4}{\ms}$ and a cut-off frequency equal to the Nyquist frequency, i.e. $f_s/2$, are typically used.

Using the Paterson approach with $T_\omega=\infty$ is equivalent to compute \eqref{eq:ISM} in the frequency domain as the sum of complex exponential functions as proposed in \cite{radlovic_equalization_2000} \cite{antonio_reverberation_2002}, but using shorter window lengths reduces the computational complexity of the algorithm.

\subsubsection{Negative reflection coefficients}

Using positive reflection coefficients as proposed in \cite{allen_image_1979} generates a low frequency artifact that must be removed using a high-pass filter. In addition, while a RIR recorded in a real room has both positive and negative peaks, all peaks generated by the ISM are positive. Using negative reflection coefficients as proposed in \cite{antonio_reverberation_2002} solve both problems without the need for adding any posterior filter to the ISM algorithm.

\subsubsection{Diffuse reverberation}

In order to properly simulate a RIR, we need to use values of $N_x$, $N_y$ and $N_z$ high enough to get all the reflections which arrive in the desired reverberation time. Since the delays of the signals of each image source are proportional to their distance to the receiver, and the distance is to the image index, the number of images to calculate for each dimension grows linearly with the reverberation time, and, therefore, the number of operations in \eqref{eq:ISM} grows in a cubic way.

A popular solution to allow the simulation of long reverberation times in a reasonable time is decomposing the RIR in two parts: the early reflections and the late, or diffuse, reverberation. While the early reflections need to be correctly simulated with the ISM method to avoid loosing spatial information, the diffuse reverberation can be modeled as a noise tail with the correct power envelope. In \cite{lehmann_diffuse_2010}, Lehmann and Johansson propose using noise with logistic distribution and the technique introduced in \cite{lehmann_prediction_2008} to predict the power envelope.

Although the technique presented in \cite{lehmann_prediction_2008} generates better predictions of the power envelope obtained in real rooms, its computational complexity is quite high. Therefore, for the sake of computational efficiency, we decided to use a simple exponential envelope following the popular Sabine formula \cite{sabine_collected_1922}. According to this model, the reverberation time $T_{60}$ that takes for a sound to decay by \SI{60}{\dB} in a room, is
\begin{equation}
    T_{60} = \frac{0.161 V}{\sum{S_i \alpha_i}},
\end{equation}
where $V$ is the volume of the room and $S_i$ and $\alpha_i=1-\beta_i^2$ are the surface area and the absorption coefficient of each wall\footnote{It should be noted that, as done in \cite{allen_image_1979}, we are defining the absorption ratio $\alpha$ as a quotient of sound intensities (energies) while the reflection coefficient $\beta$ is defined as a quotient of pressures (amplitudes).}; and the power envelope of the RIR is
\begin{equation}
    P(t) = \begin{cases}
        A \exp{\left(\log_{10}\left(\frac{T_{60}}{20}\right)(t-t_0)\right)} & \text{if } t>t_0 \\
        0 & \text{otherwise}
    \end{cases}.
\end{equation}

Therefore, knowing $T_{60}$, we can easily estimate $A$ from the early reflections simulated with the ISM and then multiply the logistic-distributed noise by $\sqrt{P(t)}$ to simulate the diffuse reverberation.

\section{Parallel implementation}
\label{sec:implementation}

\begin{figure}[]
    \centering
    \includegraphics[width=0.75\textwidth]{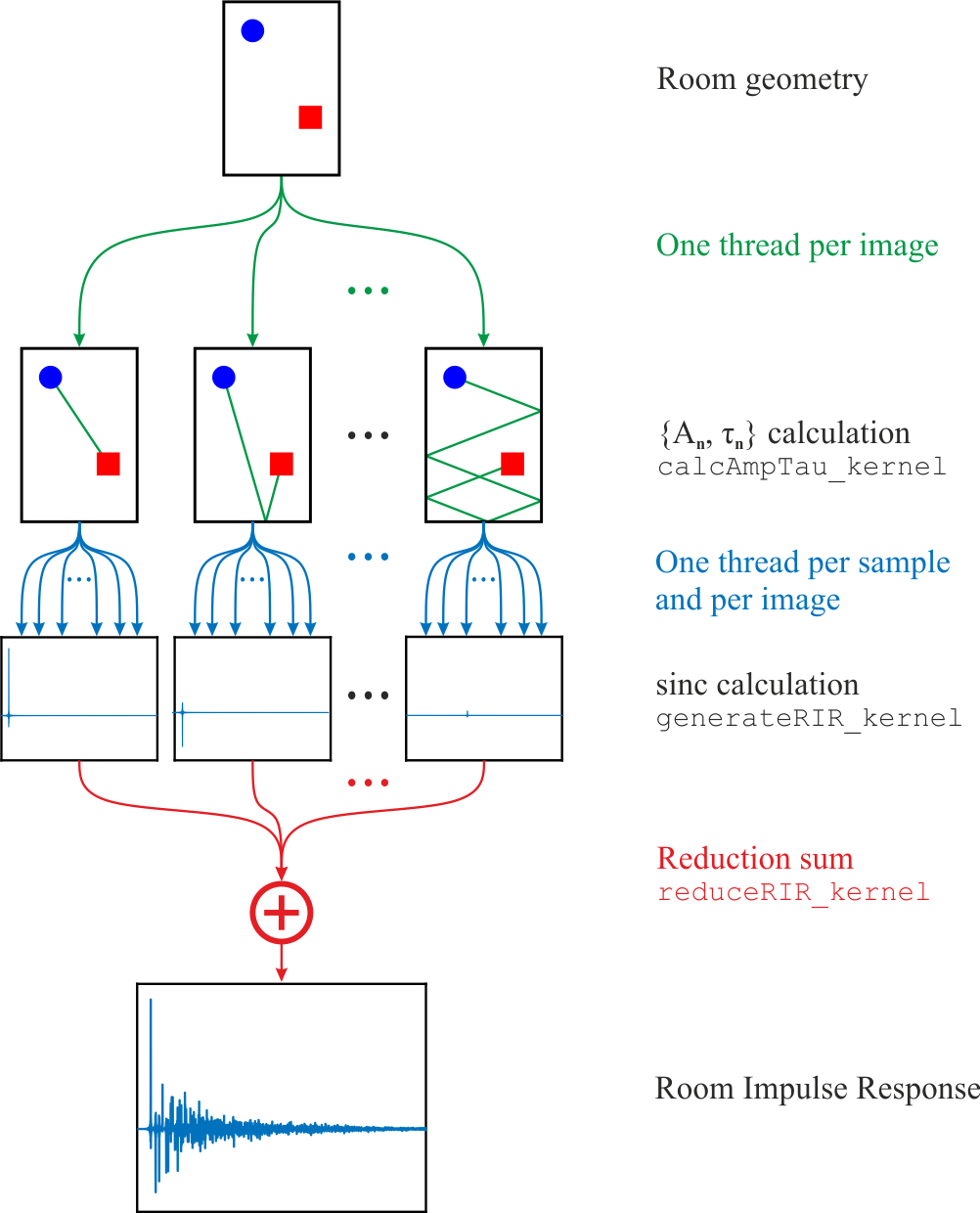}
    \caption{ISM parallel implementation. Our library actually computes some of the sincs sequentially, which leads to a more efficient memory use. The reduction sum is detailed in Fig.\ref{fig:reduction}}
    \label{fig:parallel}
\end{figure}

As shown in Fig.\ref{fig:parallel}, the parallel computation of the delays and the amplitudes of arrival for the signals from each image source and their sinc functions is straightforward since there are not any dependencies between each image source, and computing RIRs for different source or receiver positions in parallel is also trivial. However, the parallelization of \eqref{eq:ISM} involves more problems, as the contributions of all the image sources need to be added to the same RIR. 

It is worth mentioning that, though it would be possible to compute RIRs from different rooms in parallel, we choose to implement only the parallelization of RIRs corresponding to the same room. This was because the number of image sources to be computed depends on the room dimensions and the reverberation time and to compute different rooms in parallel we would have needed to use the worst case scenario (i.e. the smallest room and higher reverberation time) for all of them, which would have decreased the average performance.

%\begin{table}[]
%\footnotesize
%\centering
%    \begin{tabular}{@{}l@{\hskip5pt}l@{\hskip5pt}l@{}}
%        \toprule
%        \small{CUDA functions}      & \small{Description}      %                         & \small{Time (\%)} \\ \midrule
%        calcAmpTau\_kernel          & Equations \eqref{eq:tau} and \eqref{eq:amp}       & \small{0.82}\%    \\
%        generateTime\_kernel        & Time vector computation  %                         & \small{0.00}\%    \\
%        generateRIR\_kernel         & Sincs computation and initial sum \eqref{eq:ISM} & \small{90.1}4\%   \\
%        reduceRIR\_kernel           & Parallel sum \eqref{eq:ISM}                      & \small{1.02}\%    \\
%        generate\_seed\_pseudo      & cuRAND function (diffuse reverberation)           & \small{7.88}\%    \\
%        gen\_sequenced              & cuRAND function (diffuse reverberation)           & \small{0.01}\%    \\
%        envPred\_kernel             & Power envelope prediction %                        & \small{0.03}\%    \\
%        diffRev\_kernel             & Diffuse reverberation computation                 & \small{0.03}\%    \\
%        CUDA memcpy                 & {[}CPU to GPU{]}                                  & \small{0.00}\%    \\
%        CUDA memcpy                 & {[}GPU to CPU{]}                                  & \small{0.06}\%    \\ 
%        \bottomrule
%    \end{tabular}
%\caption{Kernels and functions of the CUDA implementation}
%\label{tab:kernels}
%\end{table}

\begin{table}%[tbp]
\caption{Kernels and functions of the CUDA implementation}
\label{tab:kernels}
\begin{tabular}{llr}
    \hline\noalign{\smallskip}
    CUDA functions              & Description                                       & Time (\%) \\
    \noalign{\smallskip}\hline\noalign{\smallskip}
    calcAmpTau\_kernel          & Equations \eqref{eq:tau} and \eqref{eq:amp}       & 0.68    \\
    generateRIR\_kernel         & Sincs computation and initial sum \eqref{eq:ISM}  & 90.34   \\
    reduceRIR\_kernel           & Parallel sum \eqref{eq:ISM}                       & 1.07    \\
    envPred\_kernel             & Power envelope prediction                         & 0.03    \\
    generate\_seed\_pseudo      & cuRAND function (diffuse reverberation)           & 7.78    \\
    gen\_sequenced              & cuRAND function (diffuse reverberation)           & 0.01    \\
    diffRev\_kernel             & Diffuse reverberation computation                 & 0.01    \\
    CUDA memcpy                 & {[}CPU to GPU{]}                                  & 0.00    \\
    CUDA memcpy                 & {[}GPU to CPU{]}                                  & 0.06    \\ 
    \noalign{\smallskip}\hline
\end{tabular}
\end{table}

In order to implement the ISM in GPUs, we decided to use CUDA \cite{nickolls_scalable_2008} and divide our code into the kernels\footnote{A CUDA kernel is a function that, when is called, is executed N times in parallel by N different CUDA threads in the GPU. For more details, see the CUDA programming guide: \url{https://docs.nvidia.com/cuda/cuda-c-programming-guide/}} listed in Table \ref{tab:kernels}. For illustrative purposes, we show in Table \ref{tab:kernels} the average proportion of time employed by each kernel to compute a standard case of 6 RIRs with $T_{60}=\SI{1}{\s}$ using the ISM method for the 250 first milliseconds and the diffuse model for the following 750ms using a Nvidia\texttrademark{} GTX 980Ti. It can be seen how the bottleneck is located at the beginning of the computation of \eqref{eq:ISM}, which is due to the high amount of sinc functions that are needed to be computed. The following sections provide further details about the implementation of the different parts of the algorithm.

\subsection{Amplitudes and delays computation}

%For computing \eqref{eq:tau} and \eqref{eq:amp} for each image source in each pair of source and receiver positions we decided to use a 3D grid of CUDA thread groups each one with 4 threads in each dimension (the number of threads per thread group, in this and the following kernels, were chosen empirically to optimize the computational performance). Each one of this threads compute sequentially the amplitudes and the delays for each RIR for one image source. The source and reciver positions are copied to the shared memoy fo the thread groups before the computations in order to redcue the number of access to global memory and increase the performance of the kernel.

For computing \eqref{eq:tau} and \eqref{eq:amp}, we use \texttt{calcAmpTau\_kernel}, which computes sequentially each RIR but parallelizes the computation for each image source. Although parallelizing the computations for each RIR would have been possible, since $N_x \cdot N_y\cdot N_z$ is generally greater than the number of RIRs to compute, the level of parallelization is already quite high and, as shown in Table \ref{tab:kernels}, further optimizations of this kernel would have had a slight impact on the final performance of the simulation.

\subsection{Computation and sum of the contribution of each image source} % Titulo demasiado largo

The computation of \eqref{eq:ISM} is the most complex part of the implementation as it implies a reduction operation (the sum of the contributions of each image source into the final RIR), which is hard to parallelize since it would imply several threads writing in the same memory address, and the calculation of a high number of trigonometric functions. We can see it as creating a tensor with 3 axis (each RIR, each image source, and each time sample) and summing it along the image sources axis. However, the size of this tensor would be huge and it would not fit in the memory of most GPUs. 

To solve this problem, we first compute and sum a fraction of the sources contributions sequentially, so the size of the tensor we need to allocate in the GPU memory is reduced; we do that through \texttt{generateRIR\_kernel}. Specifically, each parallel thread of this kernel performs sequentially the sum of 512 images for a time sample of a RIR. This sequential sum reduces the degree of parallelism of the implementation but, since the number of threads is already high enough to keep the GPU always busy, it does not decrease the performance. It should be noted that, although all the threads can potentially run in parallel, the number of threads which actually run in parallel is limited by the number of CUDA cores of the GPU and, if we have more threads than CUDA cores, many threads will be queued and will run sequentially.

\begin{figure}[]
    \centering
    \includegraphics[width=0.75\textwidth]{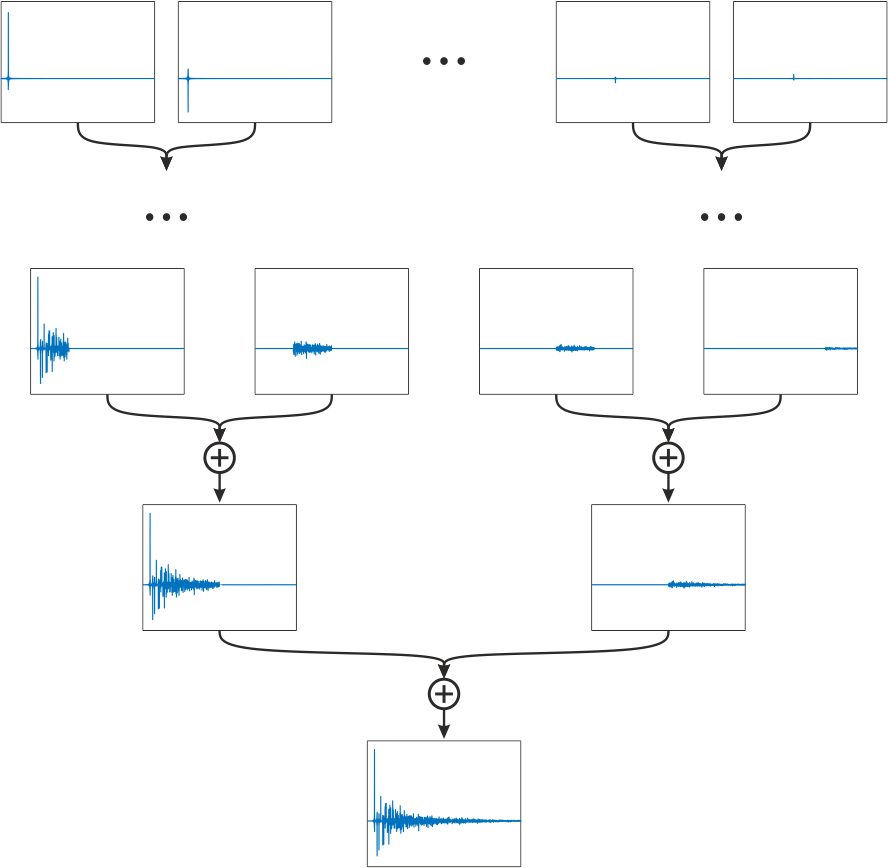}
    \caption{Parallel reduction sum of the sincs (each level is performed by a call to \texttt{reduceRIR\_kernel}). The sum must be performed pairwise to avoid several threads to concurrently write in the same variable. The sums of each time sample are also performed in parallel.}
    \label{fig:reduction}
\end{figure}

After that, we use \texttt{reduceRIR\_kernel} recursively to perform the reduction in parallel by pairwise summing the contribution of each group of images as shown in Fig.\ref{fig:reduction}. Performing the whole sum in parallel would lead to all the threads concurrently writing in the same memory positions, which would corrupt the result.

It can be seen in Table \ref{tab:kernels} how most of the simulation time is expended in \texttt{generateRIR\_kernel}, this is due to the high amount of sinc functions that need to be computed and it also happens in the sequential implementations. However, thanks to the computing power of modern GPUs, we can compute many sinc functions in parallel and therefore reduce the time we would have needed to sequentially compute them in a CPU. We analyze the implementation of these sinc functions using lookup tables (LUTs) in section \ref{sec:lut} and its performance in section \ref{sec:lut_performance}.

\subsection{Diffuse reverberation computation}

For the diffuse reverberation, we first use \texttt{envPred\_kernel} to predict in parallel the amplitude and the time constant of each RIR. After that, we use the cuRAND library included in the CUDA Toolkit to generate a uniformly distributed noise (the functions \texttt{generate\_seed\_pseudo} and \texttt{gen\_sequenced} in Table \ref{tab:kernels} belong to this library) and we finally transform it to a logistic distributed noise and apply the power envelope through \texttt{diffRev\_kernel}, which parallelizes the computations of each sample of each RIR. The function \texttt{generate\_seed\_pseudo} generates the seed for the cuRAND random number generator and it is only called when the library is imported, not every time a new RIR is calculated.

\subsection{Simulating moving sources}

As an application example of the library, it is possible to simulate a moving source recorded by a microphone array. In this case, we would need to compute the RIR between each point of the trajectory and each microphone of the array and filter the sound source by them using the overlap-add method. In sequential libraries, the complexity of the filtering is negligible compared to the RIR simulation; however, in our library, thanks to the performance of the GPUs, we found that we also needed to parallelize the filtering process if we did not want to be limited by it (specially for short reverberation times). To solve this problem, our library is able to compute multiple convolutions in parallel using the cuFFT library (included in the CUDA Toolkit) and a custom CUDA kernel to perform the pointwise complex multiplication of the FFTs.

\subsection{Lookup Tables (LUTs)}
\label{sec:lut}

Motivated by the performance increase that the CPU implementations achieve by using lookup tables (LUTs) to calculate the sinc functions (see section \ref{sec:results}), we also implemented it in our GPU library. 

Our LUT stores the values of a sinc oversampled in a factor $Q=16$ multiplied by a Hanning window:
\begin{equation}
    LUT[n] = \frac{1}{2} \left(1+\cos{\frac{2\pi n}{Q T_\omega}}\right) \text{sinc}\left(\pi \frac{n}{Q}\right) \;\; \text{for} \; n \in \left\{\frac{-T_\omega}{2}Q f_s, ... , \frac{T_\omega}{2}Q f_s\right\}
\end{equation}
% \begin{equation}
%     LUT[n] = \begin{cases} \frac{1}{2} \left(1+\cos{\frac{2\pi n}{Q T_\omega}}\right) \text{sinc}(\pi \frac{n}{Q}) & \text{if } n \in \{\frac{-T_\omega}{2}Q f_s, ... , \frac{T_\omega}{2}Q f_s\} \setminus \{0\} \\
%     1.0 & \text{if } n = 0
%     \end{cases}
% \end{equation}
%
and then we use linear interpolation between the closest entries of the table to compute each sample of the sinc functions of each image source.

The main design choice we must make is to define the type of memory that will be used to place the LUT. CUDA GPUs have, in addition to the registers of each thread, 4 different memories: shared, global, constant and texture memory. On the one hand, shared memory is shared only between threads of the same block and it has the fastest access, however it is generally lower than 100KB. On the other hand, global memory is shared by all the threads and usually has several gigabytes, but it has the lower bandwidth and the higher latency. Finally, constant and texture memories are read-only cached memories, constant memory being optimized for several threads accessing to the same address and texture memory being optimized for memory access with spatial locality. Although constant memory has a lower latency than texture memory, texture memory implements some features like several accessing modes and hardware interpolation, which are extremely useful for the implementation of LUTs. We implemented the windowed sinc LUT both in shared memory and texture memory and obtained better performance with the texture memory thanks to the hardware interpolation.

\subsection{Mixed precision}

Since the Pascal architecture, the Nvidia\textsuperscript{TM} GPUs include support for 16-bit precision floats and are able to perform two 16 bit operations at a time. To exploit this feature, we developed the kernels \texttt{generateRIR\_mp\_kernel} and \texttt{reduceRIR\_mp\_kernel}, which compute two consecutive time samples at a time so we can halve the number of threads needed. We focused on these kernels and did not optimise the others because, as shown in Table \ref{tab:kernels}, most of the simulation time is spent in them.

CUDA provides the data type \texttt{half2}, which contains 2 floating point numbers of 16 bits, and several intrinsics to operate with it. These intrinsics allow to double the number of arithmetic operations that we can perform per second; however, we found that the functions provided to compute two 16-bit trigonometric functions were not as fast as computing one 32-bit function. To increase the simulation speed, we developed our own \texttt{sinpi(half2)} and \texttt{cospi(half2)} functions. For the sine function we first reduce the argument to the range [-0.5, 0.5], then we approximate the sine function in this range by
\begin{equation}
    \sin(\pi x) \approx 2.326171875x^5 - 5.14453125x^3 + 3.140625x
\end{equation}
and finally, multiply the result by -1 if the angle was in the second or the third quadrant. The coefficients of the polynomial are the closest numbers that can be represented with half precision floats to those of the optimal polynomial in a least-squares sense. Equivalently, for the cosine function, we used the polynomial:
\begin{equation}
    \cos(\pi x) \approx -1.2294921875x^6 + 4.04296875x^4 - 4.93359375x^2 + 1
\end{equation}
with the advantage that, since we only used it for computing the Hanning window in (\ref{eq:ISM}), we do not need to perform argument reduction or sign correction. 

The polynomial evaluation can be efficiently performed with the Horner's method:
\begin{equation}
\begin{split}
    &b_n = a_n \\
    &b_{n-1} = a_{n-1} + b_{n} x \\
    &... \\
    &p(x) = b_0 = a_0 + b_1 x
\end{split}
\end{equation}
where $a_i$ are the coefficient of the $n$ degree polynomial $p(x)$ we want to evaluate and the computation of $b_i$ can be done in parallel for two different values of $x$ using the CUDA intrinsic \texttt{\_\_hfma2(half2)} that performs the fused multiply-add operation of the two elements of three \texttt{half2} variables at a time. More information about polynomial approximation of transcendental functions can be found in \cite{myklebust_computing_2015}.

Obviously, working with half precision representation reduces the accuracy of the results. We found that the most critical part was in subtracting $t-\tau_n$. Working with 16-bit precision floats, we can only represent 3 significant figures accurately, so, when $t$ grows, we lose precision in the argument of the sinc function which leads to an error which increases with the time; when $t$ grows we expend the precision in the integer part and we don't represent accurately the fractional part. To solve this issue, we perform the subtraction with 32 bits arithmetic and then we transform the result to 16-bit precision. Working this way, we have always maximum precision in the centre of the sinc and the lower accuracy is outside the Hanning window.

Unfortunately, the hardware interpolation of the texture memory does not support 16-bit arithmetic, so the mixed precision implementation is not compatible with the LUT. 

\section{Python library}
\label{sec:library}

We have included the previous implementation in a Python library that can be easily compiled and installed using the Python packet manager (pip) and be used as any CPU library. The library provides a function which takes as parameters the room dimensions, the reflections coefficients of the walls, the position of the source and the receivers, the number of images to simulate for each dimension, the duration of the RIR in seconds, the time to switch from the ISM method to the diffuse reverberation model, and the sampling frequency and it returns a 3D tensor with the RIR for each pair of source and receiver positions. Information about the polar pattern of the receivers and their orientation can be also included in the simulation.

We also provide some python functions to predict the time when some level of attenuation will be reached, to get the reflections coefficients needed to get the desired reverberation time (expressed in terms of $T_{60}$, i.e. the time needed to get an attenuation of \SI{60}{\dB}), and to get the number of image sources to simulate in each dimension to get the desired simulation time without loss reflections. Finally, we include a function to filter a sound signal by several RIRs in order to simulate a moving source recorded by a microphone array. In the repository of the library some examples can be found about how to simulate both isolated RIRs and moving sources.

Since the use of the LUT to compute the sinc function improves the performance in most of the cases and the precision loss is negligible (see section \ref{sec:lut_performance}), its use is activated by default, but the library provides a function to deactivate it and use the CUDA trigonometric functions instead. In order to exploit the mixed precision capabilities of the newer GPUs, it has a function to activate it and use the 16-bit precision kernels instead of the 32-bit; activating it automatically deactivates the use of the LUT. 

Since the library was developed, we have used it to train a sound source tracking system based on a 3D Convolutional Neural Network simulating the training signals as they were needed instead of creating a pre-simulated dataset \cite{diaz-guerra_robust_2020}; this approach has the advantage of being equivalent to have an infinite-size dataset, but it would have been unfeasible with the simulation times of previous libraries. Other authors have also used it to train deep learning systems \cite{luo_fasnet_2019,luo_end--end_2020,wang_neural_2020,mirbagheri_c-sl_2020} and to evaluate signal processing techniques \cite{ceolini_evaluating_2020,ziegler_acoustic_2020}.

\section{Results}
\label{sec:results}

\subsection{Base implementation}

In order to show the benefits of using GPUs for RIR simulation, we have compared our library against three of the most employed libraries for this purpose: the Python version of the RIR Generator library presented in \cite{habets_room_2010}, whose code is freely available in \cite{marvin182_room_2018} and has been used, for example, in \cite{williamson_time_2017,hassani_multi-task_2017,markovich_multichannel_2009}; the Python package pyroomacoustics presented in \cite{scheibler_pyroomacoustics_2018} that has been employed in \cite{qin_far-field_2019,mosner_improving_2019,severini_automatic_2019} among others; and the Matlab\texttrademark{} library presented in \cite{lehmann_diffuse_2010}, whose code is freely available in \cite{lehmann_matlab_2018}, and that has been used, for example, in \cite{griffin_localizing_2015,pavlidi_real-time_2012,alexandridis_capturing_2013}. Since all the libraries are based on the ISM, whose acoustical accuracy is well known, we focus on the computation time of each library.

Neither RIR Generator nor pyroomacoustics implement any kind of diffuse reverberation model, so they are expected to have worse performance than the Matlab\texttrademark{} library and our GPU library if we use it. The Matlab\texttrademark{} library uses the formula presented in \cite{lehmann_prediction_2008} to model the power envelope of the diffuse reverberation, which is more complex than our exponential envelope model, so, for the sake of a fairer comparison, we modified the Matlab\texttrademark{} implementation to use a exponential model. The simulations with the sequential libraries and the ones with the Nvidia\texttrademark{} GTX 980Ti were performed in a computer with an Intel\texttrademark{} Core i7-6700 CPU and \SI{16}{\giga\byte} of RAM, while the simulations with the Nvidia\texttrademark{} Tesla P100 and V100 and T4 were performed in a n1-highmem-4 instance in the Google Cloud Platform\texttrademark{} with 4 virtual CPUs cores and \SI{26}{\giga\byte} of RAM memory; more details about the GPUs employed for the simulations can be found in Table \ref{tab:gpus}.

\begin{table}%[tbp]
\caption{GPUs employed for the performance analysis}
\label{tab:gpus}
\begin{tabular}{lllll}
    \hline\noalign{\smallskip}
    GPU model       & Architecture  & Memory    & Single Precision FLOP/s   & Memory Bandwidth\\
    \noalign{\smallskip}\hline\noalign{\smallskip}
    GTX 980 Ti      & Maxwell       & 6GB       & 5.6 TeraFLOP/s            & 337 GB/s \\
    Tesla P100      & Pascal        & 16GB      & 9.5 TeraFLOP/s            & 732 GB/s \\
    Tesla V100      & Volta         & 16GB      & 14.9 TeraFLOP/s           & 900 GB/s \\
    Tesla T4        & Turing        & 16GB      & 8.1 TeraFLOP/s            & 320 GB/s \\
    \noalign{\smallskip}\hline
\end{tabular}
\end{table}

\begin{figure}[]
    \centering
    \includegraphics[width=0.75\textwidth]{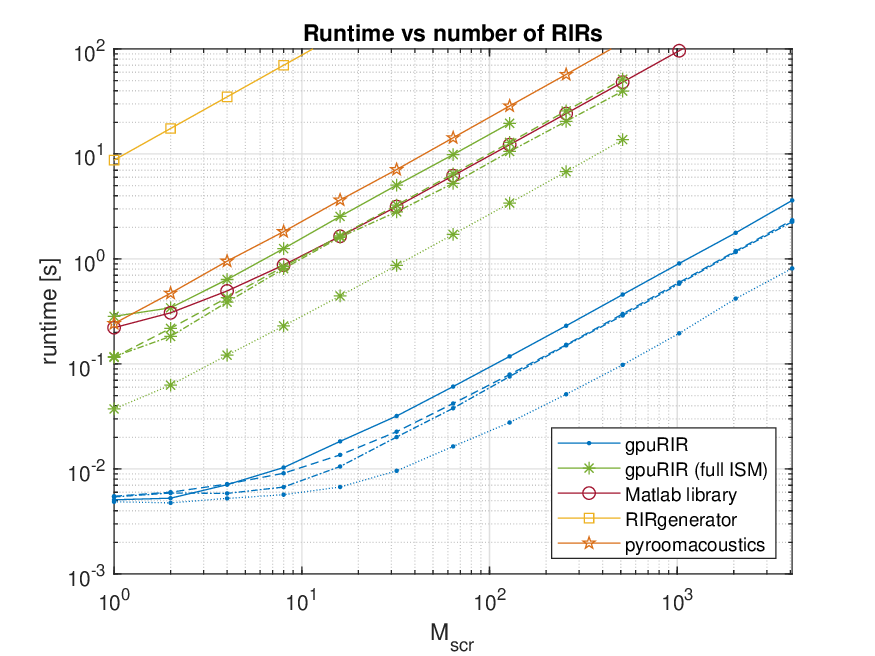}
    \caption{Runtime of each library for computing different numbers of RIRs ($\mathrm{M_{src}}$) in a room with size \SI{3x4x2.5}{\m} and $T_{60}=\SI{0.7}{\s}$. For the gpuRIR library, the solid line times were obtained with the GTX 980 Ti GPU, the dashed lines with the Tesla P100, the dotted lines with the Tesla V100, and the dash-dot lines with the Tesla T4.}
    \label{fig:runtime_Msrc}
\end{figure}

Fig.\ref{fig:runtime_Msrc} represents the runtime of the different libraries for computing different numbers of RIRs in a room with size \SI{3x4x2.5}{\m} and $T_{60}=\SI{0.7}{\s}$. It can be seen how our library can simulate a hundred times more RIRs in a second than the Matlab\texttrademark{} library even with a GPU designed for gaming (the Nvidia\texttrademark{} GTX 980 Ti). Using our library without any kind of diffuse reverberation modeling, we have a similar execution time than the Matlab\texttrademark{} library, which only computes the ISM until the RIR has an attenuation of \SI{13}{\dB}, and we are also about a hundred times faster than the RIR Generator library. Finally, it is worth noting how pyroomacoustics performs quite similarly to our library when we use a GTX 980 Ti and compute the whole RIR with the ISM without using any diffuse reverberation model; this is due to the use of LUTs to compute the sinc functions by pyroomacoustics (to confirm this hypothesis we modified the code of pyroomacoustics to avoid the use of LUTs and its performance degraded to the same results than RIR Generator). However, using a faster GPU, i.e. the Tesla V100, our library can compute ten times more RIRs in a second than pyroomacoustics even without using LUTs, since we can set at full performance all the parallelization mechanisms presented in section \ref{sec:implementation}.

Comparing the performance of our library using different GPUs, we can see how the lower results are obtained using the GTX 980 Ti, the Tesla P100 and T4 have a quite similar performance (being the T4 slightly faster), and the better results are obtained with the Tesla V100 (being more than 5 times faster than the GTX 980 Ti). This results are what we could expect for an algorithm whose computation time is mostly limited by the number of operations that we can perform per second, but it is worth noting how the Tesla T4 (with the newer Nvidia\texttrademark{} GPU architecture) can outperform the Tesla P100 having lower FLOP/s, memory bandwidth and power consumption.

\begin{figure}[]
    \centering
    \includegraphics[width=0.75\textwidth]{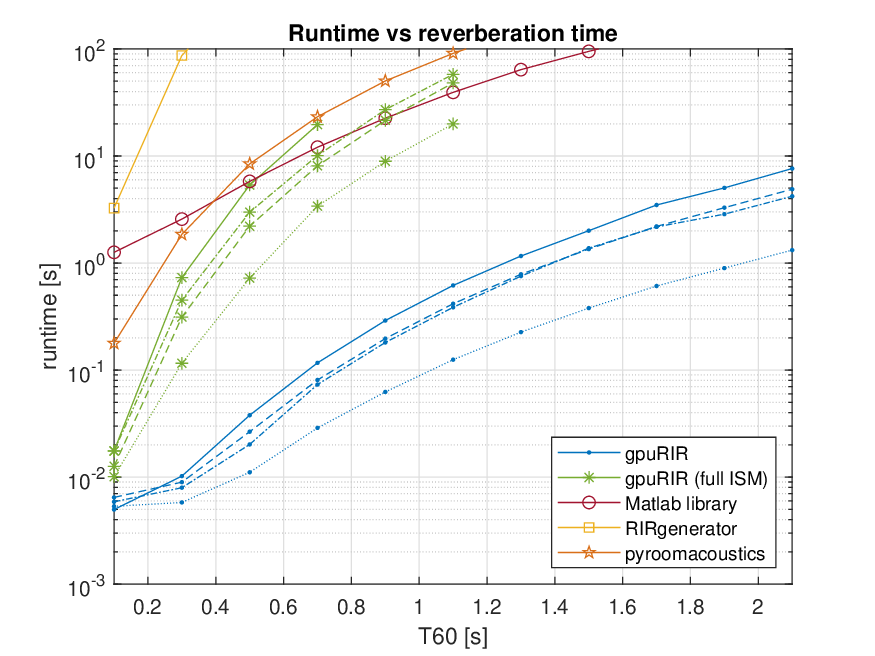}
    \caption{Runtime of each library for computing 128 RIRs in a room with size \SI{3x4x2.5}{\m} and different reverberation times. For the gpuRIR library, the solid line times were obtained with the GTX 980 Ti GPU, the dashed lines with the Tesla P100, the dotted lines with the Tesla V100, and the dash-dot lines with the Tesla T4.}
    \label{fig:runtime_T60}
\end{figure}

In Fig.\ref{fig:runtime_T60} we show the runtime of the different libraries for computing 128 RIRs in a room with size 3x4x2.5m and different reverberation times. We can see again how our library is about two orders of magnitude faster than the sequential alternatives which do not use LUTs. It must be said that our library has some limitations because calculating a large number of RIRs with high reverberation times may require more memory than it is available in the GPU; however, using the diffuse reverberation model, this limitation appears only for really high number of RIRs and reverberation times. Furthermore, it would be always possible to batch the RIRs in several function calls to circumvent this problem.

%We did not perform any simulation campaign to analyze the acoustic accuracy of the library as it is the same of the well known ISM method, but we have compared our results against the cited libraries to validate it.

\subsection{Lookup tables}
\label{sec:lut_performance}

Motivated by the huge speedup generated by the use of LUTs in the CPU implementations (a factor 5 in Fig. \ref{fig:runtime_Msrc}) we replaced the trigonometric computations by a LUT as described in section \ref{sec:lut}. Tables \ref{tab:speedupM} and \ref{tab:speedupT} show the speedup (defined as the runtime without using the LUT divided by the runtime using it) for several numbers of RIRs and reverberation times using different GPUs. 

We can see how our library obtains a speedup much lower than the obtained by pyroomacoustics over CPU. This is due to the high computation power of the GPUs, which makes the computation of trigonometric functions quite efficient and therefore they are not so benefited by replacing computation tasks by memory calls. Despite that, we can see how using LUTs is faster than computing the trigonometric functions, i.e. the speedup is higher than 1.0, in most of the cases, especially when the number of RIRs or the reverberation time increases.

\begin{table}%[tbp] 
\renewcommand{\arraystretch}{1.3}
\caption{Lookup Table (LUT) and Mixed Precision (MP) simulation times and speedups for computing different numbers of RIRs with $T_{60}=\SI{0.7}{\s}$}
\label{tab:speedupM}
\begin{tabular}{lM{12.5mm}M{8mm}M{9mm}M{10mm}M{10mm}cM{10mm}M{11mm}M{12mm}}
    \toprule%\hline\noalign{\smallskip}
    \multicolumn{2}{r}{\multirow{2}{*}{Number of RIRs}} & \multicolumn{4}{c}{Diffuse reverberation model} & \phantom{} & \multicolumn{3}{c}{Full ISM} \\
    \cmidrule{3-6} \cmidrule{8-10} 
    &                          & 1                  & 16                  & 128                  & 1024                 && 1                    & 16                      & 128                    \\
    \midrule%\noalign{\smallskip}\hline\noalign{\smallskip}
    \multicolumn{2}{l}{Matlab Library}
                               & 221,52             & 1,643.20            & 12,252.67            & 96,208.58            && -                    & -                       & -                      \\
    \multicolumn{2}{l}{pyroomacoustics}
                               & -                  & -                   & -                    & -                    && 242.35               & 3,6409.16               & 28,646.86              \\
    \midrule%\noalign{\smallskip}\hline\noalign{\smallskip}
    \multirow{3}{*}[-1.75mm]{\rotatebox[origin=c]{90}{GTX 980 Ti}}
    & Base [ms]                & 4.98               & 17.43               & 117.60               & 898.54               && 283.88               & 2,601.82               & 19,630.60               \\
    & LUT [ms] \light{speedup} & 5.19 \light{x0.96} & 16.64 \light{x1.05} & 109.38 \light{x1.08} & 834.38 \light{x1.08} && 279.28 \light{x1.02} & 2,434.33 \light{x1.07} & 18,547.03 \light{x1.06} \\
    & MP [ms]  \light{speedup} & -                  & -                   & -                    & -                    && -                    & -                      & -                       \\
    \midrule%\noalign{\smallskip}\hline\noalign{\smallskip}
    \multirow{3}{*}[-2.5mm]{\rotatebox[origin=c]{90}{Tesla P100}}
    & Base [ms]                & 5.81               & 13.86               & 79.28                & 596.02               && 115.5       7        & 1,661.35               & 12,879.31               \\
    & LUT [ms] \light{speedup} & 5.97 \light{x0.97} & 12.14 \light{x1.14} & 63.90 \light{x1.24}  & 471.16 \light{x1.27} && 86.86 \light{x1.33}  & 1,235.64 \light{x1.35} & 9,397.40 \light{x1.37}  \\
    & MP [ms]  \light{speedup} & 5.52 \light{x1.05} & 9.45 \light{x1.47}  & 45.49 \light{x1.74}  & 324.12 \light{x1.84} && 59.46 \light{x1.94}  & 847.74 \light{x1.96}   & 6,493.92 \light{x1.98}  \\
    \midrule%\noalign{\smallskip}\hline\noalign{\smallskip}
    \multirow{3}{*}[-2.5mm]{\rotatebox[origin=c]{90}{Tesla V100}}
    & Base [ms]                & 4.76               & 7.13                & 28.14                & 195.69               && 37.62                & 447.04                 & 3,403.60                \\
    & LUT [ms] \light{speedup} & 5.01 \light{x0.95} & 6.79 \light{x1.05}  & 23.66 \light{x1.19}  & 156.91 \light{x1.25} && 30.66 \light{x1.23}  & 394.54 \light{x1.13}   & 2,595.97 \light{x1.31}  \\
    & MP [ms]  \light{speedup} & 4.55 \light{x1.05} & 6.29 \light{x1.13}  & 19.57 \light{x1.44}  & 128.72 \light{x1.52} && 21.76 \light{x1.73}  & 253.03 \light{x1.77}   & 1,900.52 \light{x1.79}  \\
    \midrule%\noalign{\smallskip}\hline\noalign{\smallskip}
    \multirow{3}{*}[-3mm]{\rotatebox[origin=c]{90}{Tesla T4}}
    & Base [ms]                & 5.80               & 10.95               & 73.49                & 582.79               && 117.00               & 1,612.79               & 10,188.68               \\
    & LUT [ms] \light{speedup} & 5.63 \light{x1.03} & 10.14 \light{x1.08} & 63.75 \light{x1.15}  & 503.91 \light{x1.16} && 81.37 \light{x1.44}  & 1,433.60 \light{x1.13} & 8,870.68 \light{x1.15}  \\
    & MP [ms]  \light{speedup} & 4.80 \light{x1.21} & 7.37 \light{x1.43}  & 43.28 \light{x1.76}  & 351.78 \light{x1.66} && 58.45 \light{x2.00}  & 860.43 \light{x1.87}   & 5,693.29 \light{x1.79}  \\
    \bottomrule
\end{tabular}
\end{table}

\begin{table}%[tbp] 
\renewcommand{\arraystretch}{1.3}
\caption{Lookup Table (LUT) and Mixed Precision (MP) simulation times and speedups for computing different numbers of RIRs with $T_{60}=\SI{0.7}{\s}$}
\label{tab:speedupT}
\begin{tabular}{lM{12.5mm}M{11mm}M{11mm}M{11mm}M{11mm}M{12mm}cM{11mm}M{12mm}M{12mm}}
    \toprule%\hline\noalign{\smallskip}
    \multicolumn{2}{r}{\multirow{2}{*}{$T_{60}$ [s]}} & \multicolumn{5}{c}{Diffuse reverberation model} & \phantom{} & \multicolumn{3}{c}{Full ISM} \\
    \cmidrule{3-7} \cmidrule{9-11} 
    &                          & 0.3                & 0.7                  & 1.1                  & 1.5                    & 1.9                    && 0.3                  & 0.7                     & 1.1                     \\
    \midrule%\noalign{\smallskip}\hline\noalign{\smallskip}
    \multicolumn{2}{r}{Matlab Library}
                               & 2,573.67           & 12,078.52            & 39,330.40            & 94,946.73              & 136,522.39             && -                    & -                       & -                       \\
    \multicolumn{2}{r}{pyroomacoustics}
                               & -                  & -                    & -                    & -                      & -                      && 1,854.08             & 23,253.22               & 90,960.54               \\
    \midrule%\noalign{\smallskip}\hline\noalign{\smallskip}
    \multirow{3}{*}[-1.75mm]{\rotatebox[origin=c]{90}{GTX 980 Ti}}
    & Base [ms]                & 8.90               & 118.00               & 627.15               & 2,016.40               & 5,073.48               && 731.59               & 19,657.05               & -                       \\
    & LUT [ms] \light{speedup} & 8.62 \light{x1.03} & 109.89 \light{x1.07} & 588.90 \light{x1.06} & 1,896.48 \light{x1.06} & 4,769.68 \light{x1.06} && 694.10 \light{x1.05} & 18,466.15 \light{x1.06} & -                       \\
    & MP [ms] \light{speedup}  & -                  & -                    & -                    & -                      & -                      && -                    & -                       & -                       \\
    \midrule%\noalign{\smallskip}\hline\noalign{\smallskip}
    \multirow{3}{*}[-2.5mm]{\rotatebox[origin=c]{90}{Tesla P100}}
    & Base [ms]                & 8.97               & 80.78                & 416.57               & 1,349.47               & 3,289.13               && 494.33               & 12,875.14               & 76,383.87               \\
    & LUT [ms] \light{speedup} & 7.39 \light{x1.21} & 64.90 \light{x1.25}  & 321.97 \light{x1.29} & 1,023.39 \light{x1.32} & 2,452.18 \light{x1.34} && 391.39 \light{x1.26} & 9,406.31 \light{x1.37}  & 55,402.08 \light{x1.38} \\
    & MP [ms] \light{speedup}  & 6.64 \light{x1.35} & 45.18 \light{x1.79}  & 218.18 \light{x1.91} & 699.95 \light{x1.93}   & 1,698.17 \light{x1.94} && 258.96 \light{x1.91} & 6,484.46 \light{x1.99}  & 38,393.03 \light{x1.99} \\
    \midrule%\noalign{\smallskip}\hline\noalign{\smallskip}
    \multirow{3}{*}[-2.5mm]{\rotatebox[origin=c]{90}{Tesla V100}}
    & Base [ms]                & 5.80               & 28.81                & 125.02               & 379.13                 & 896.97                 && 141.86               & 3,400.95                & 19,935.71               \\
    & LUT [ms] \light{speedup} & 5.95 \light{x0.97} & 24.43 \light{x1.18}  & 101.85 \light{x1.23} & 332.35 \light{x1.14}   & 690.02 \light{x1.30}   && 117.55 \light{x1.22} & 2,594.15 \light{x1.31}  & 15,363.05 \light{x1.30} \\
    & MP [ms]  \light{speedup} & 5.08 \light{x1.14} & 19.80 \light{x1.46}  & 76.71 \light{x1.63}  & 220.66 \light{x1.72}   & 519.46 \light{x1.73}   && 87.48 \light{x1.62}  & 1,901.21 \light{x1.79}  & 11,052.52 \light{x1.80} \\
    \midrule%\noalign{\smallskip}\hline\noalign{\smallskip}
    \multirow{3}{*}[-3mm]{\rotatebox[origin=c]{90}{Tesla T4}}
    & Base [ms]                & 6.43               & 73.22                & 385.88               & 1,376.26               & 2,862.44               && 465.76               & 10,139.45               & 57,596.94               \\
    & LUT [ms] \light{speedup} & 6.59 \light{x0.97} & 63.20 \light{x1.16}  & 344.66 \light{x1.12} & 1,122.22 \light{x1.23} & 2,406.75 \light{x1.19} && 407.88 \light{x1.14} & 8,812.73 \light{x1.15}  & 49,612.01 \light{x1.16} \\
    & MP [ms] \light{speedup}  & 5.80 \light{x1.11} & 43.93 \light{x1.66}  & 230.89 \light{x1.67} & 770.90 \light{x1.79}   & 1,841.16 \light{x1.55} && 270.88 \light{x1.72} & 5,693.18 \light{x1.78}  & 31,377.01 \light{x1.84} \\
    \bottomrule
\end{tabular}
\end{table}

Among the studied GPUs, the Tesla P100 obtains the higher speedups since it has a higher memory bandwidth compared with its computing power. The GTX 980 Ti gets really humble speedups due it low memory bandwidth and the Tesla V100, though it has the higher bandwidth, does not reach the speedups obtained by the Tesla P100 due to its huge computing power. Finally, it is interesting how the Tesla T4 obtains higher speedups than the GTX 980 Ti despite having a lower memory bandwidth; this might be due to some optimizations introduced in the newer Turing architecture.

\begin{figure}[]
    \centering
    \includegraphics[width=0.75\textwidth]{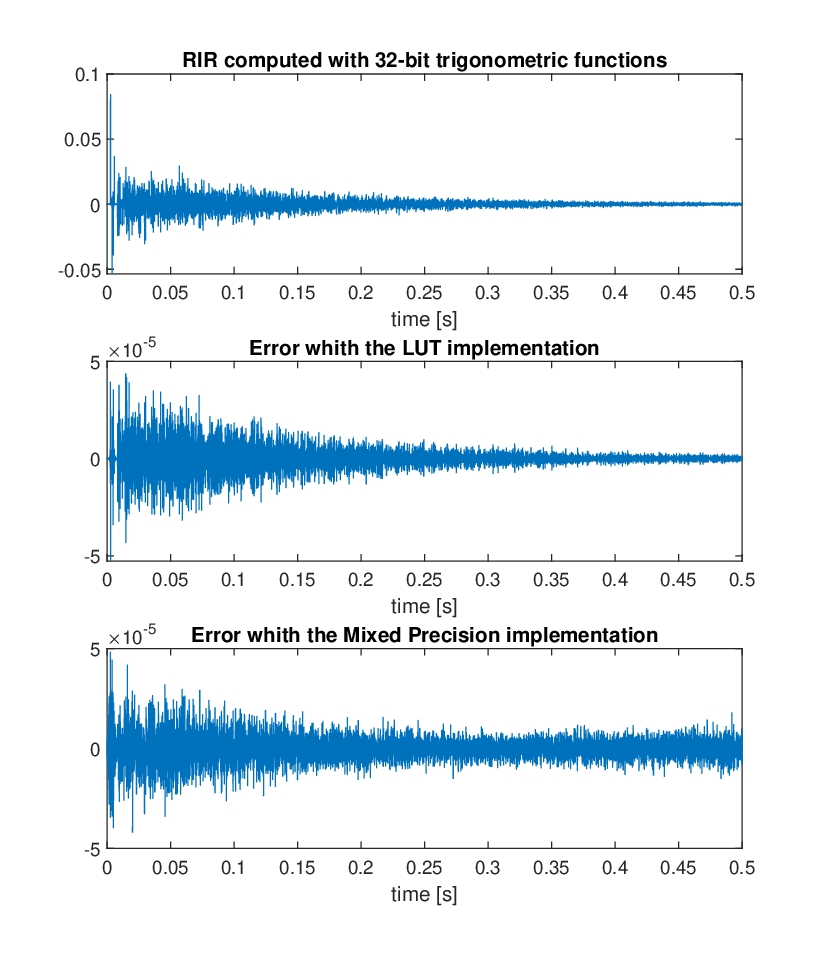}
    \caption{RIR computed with single (32-bit) precision trigonometric functions and the error introduced due to compute it using a lookup table (LUT) and half (16-bit) precision functions (Mixed Precision).}
    \label{fig:errors}
\end{figure}

Fig. \ref{fig:errors} shows the first 0.5 seconds of the RIR of a room with $T_{60}=\SI{1}{s}$ computed with our GPU implementation working with single (32-bit) precision trigonometric functions and the error introduced by replacing them by out LUT. We can see how, as it could be expected, the error introduced by the use of the LUT is negligible: three orders of magnitude lower than the amplitude of the RIR.

\subsection{Mixed precision}

In case of using the 16-bit precision kernels, we are reducing the accuracy of the simulation, so we need to analyze its impact. Fig. \ref{fig:errors} also shows the error introduced by computing the same RIR using our half (16-bit) precision kernels. We can see how the  error is 3 orders of magnitude lower than the amplitude of the RIR at the beginning, which should be acceptable for most of the applications; however, since the error does not decrease with the time as much as the RIR does, the signal-to-error ratio deteriorates with the time. Hopefully, this higher error correspond with the diffuse reverberation, where its perceptual importance is lower.

Theoretically, a twofold speedup could be expected from working with 16-bit precision floats instead of 32-bit floats, however, this speedup is generally not reachable as the number of operations is not the only limiting factor of many GPU kernels and some \texttt{half2} functions are not as fast as its equivalent \texttt{single} functions. Tables \ref{tab:speedupM} and \ref{tab:speedupT} show the speedup that our mixed precision implementation achieve for several numbers of RIRs computed in parallel and several reverberation times. We can see how the speedup is higher when the workload increases, especially for long reverberation times where the operations per second are the main limiting factor of its performance and how the speedup achieved with the mixed precision implementation is always higher than the achieved with the LUTs.

The mixed precision support was introduced with the Pascal architecture, so it is not available in older models like the GTX 980 Ti. The Tesla P100 achieves speedups really close to 2 for high workloads. The speedups obtained with the Tesla T4 are quite erratic and its increase with the workload is no so clear than with other GPUs, but it is generally higher than the speedup obtained with the Tesla V100.

\section{Conclusions}
\label{sec:conclusions}

We have presented a new free and open-source library to simulate RIRs that uses GPUs to dramatically reduce the simulation time and it has been proved that it is about one hundred times faster than other state of the art CPU libraries. To the best of out knowledge, it is the first library with these features freely available on the Internet, and it could allow to the acoustic signal processing community, for example, to generate huge datasets of moving speaker speech signals in a reasonable computation time or to compute the acoustics of a Virtual Reality (VR) scene in real time.

We have studied different methods to increase the speed of our GPU implementation, concluding that the best strategy is using 16-bit arithmetic, but this is only compatible with the newer GPUs. On the other hand, using LUTs stored in the GPU's texture memory, though it generates lower speedups, is compatible with most of the CUDA GPUs, so we have chosen to use this implementation as our library default.

We expect this library to be a useful tool for the audio signal processing community, especially for those who need to simulate large audio datasets to train their models. Since it has been published as free and open-source software, it can be easily upgraded to exploit the new features that future generations of GPUs may bring, both by us as the original authors or by any other researcher interested in it.

\begin{acknowledgements}
Authors would like to thank Norbert Juffa for his advises and assistance in the Nvidia\texttrademark{} Developer Forums.
\end{acknowledgements}

\bibliographystyle{spphys}       % APS-like style for physics
\bibliography{bibliography.bib}   % name your BibTeX data base

\begin{thebibliography}{10}
\providecommand{\url}[1]{{#1}}
\providecommand{\urlprefix}{URL }
\expandafter\ifx\csname urlstyle\endcsname\relax
  \providecommand{\doi}[1]{DOI \discretionary{}{}{}#1}\else
  \providecommand{\doi}{DOI \discretionary{}{}{}\begingroup
  \urlstyle{rm}\Url}\fi

\bibitem{weng_recurrent_2014}
C.~Weng, D.~Yu, S.~Watanabe, B.F. Juang, in \emph{2014 {{IEEE International
  Conference}} on {{Acoustics}}, {{Speech}} and {{Signal Processing}}
  ({{ICASSP}})} (2014), pp. 5532--5536.
\newblock \doi{10.1109/ICASSP.2014.6854661}

\bibitem{griffin_localizing_2015}
A.~Griffin, A.~Alexandridis, D.~Pavlidi, Y.~Mastorakis, A.~Mouchtaris,
  Localizing {{Multiple Audio Sources}} in a {{Wireless Acoustic Sensor
  Network}}, Signal Processing \textbf{107}, 54 (2015).
\newblock \doi{10.1016/j.sigpro.2014.08.013}

\bibitem{williamson_time_2017}
D.S. Williamson, D.~Wang, Time-{{Frequency Masking}} in the {{Complex Domain}}
  for {{Speech Dereverberation}} and {{Denoising}}, IEEE/ACM Transactions on
  Audio, Speech, and Language Processing \textbf{25}(7), 1492 (2017).
\newblock \doi{10.1109/TASLP.2017.2696307}

\bibitem{schroeder_natural_1962}
M.R. Schroeder, Natural {{Sounding Artificial Reverberation}}, Journal of the
  Audio Engineering Society \textbf{10}(3), 219 (1962)

\bibitem{fu_gpu-based_2016}
Z.h. Fu, J.w. Li, {{GPU}}-based image method for room impulse response
  calculation, Multimedia Tools and Applications \textbf{75}(9), 5205 (2016).
\newblock \doi{10.1007/s11042-015-2943-4}

\bibitem{habets_room_2010}
E.A. Habets, Room {{Impulse Response Generator}}.
\newblock Tech. rep. (2010)

\bibitem{scheibler_pyroomacoustics_2018}
R.~Scheibler, E.~Bezzam, I.~Dokmani{\'c}, in \emph{2018 {{IEEE International
  Conference}} on {{Acoustics}}, {{Speech}} and {{Signal Processing}}
  ({{ICASSP}})} (2018), pp. 351--355.
\newblock \doi{10.1109/ICASSP.2018.8461310}

\bibitem{lehmann_diffuse_2010}
E.A. Lehmann, A.M. Johansson, Diffuse {{Reverberation Model}} for {{Efficient
  Image}}-{{Source Simulation}} of {{Room Impulse Responses}}, IEEE
  Transactions on Audio, Speech, and Language Processing \textbf{18}(6), 1429
  (2010).
\newblock \doi{10.1109/TASL.2009.2035038}

\bibitem{allen_image_1979}
J.B. Allen, D.A. Berkley, Image method for efficiently simulating small-room
  acoustics, The Journal of the Acoustical Society of America \textbf{65}(4),
  943 (1979).
\newblock \doi{10.1121/1.382599}

\bibitem{peterson_simulating_1986}
P.M. Peterson, Simulating the response of multiple microphones to a single
  acoustic source in a reverberant room, The Journal of the Acoustical Society
  of America \textbf{80}(5), 1527 (1986).
\newblock \doi{10.1121/1.394357}

\bibitem{radlovic_equalization_2000}
B.D. Radlovic, R.C. Williamson, R.A. Kennedy, Equalization in an acoustic
  reverberant environment: Robustness results, IEEE Transactions on Speech and
  Audio Processing \textbf{8}(3), 311 (2000).
\newblock \doi{10.1109/89.841213}

\bibitem{antonio_reverberation_2002}
J.~Antonio, L.~Godinho, A.~Tadeu, Reverberation {{Times Obtained Using}} a
  {{Numerical Model Versus Those Given}} by {{Simplified Formulas}} and
  {{Measurements}}, ACTA ACUSTICA UNITED WITH ACUSTICA \textbf{88}, 10 (2002)

\bibitem{lehmann_prediction_2008}
E.A. Lehmann, A.M. Johansson, Prediction of {{Energy Decay}} in {{Room Impulse
  Responses Simulated}} with an {{Image}}-{{Source Model}}, The Journal of the
  Acoustical Society of America \textbf{124}(1), 269 (2008).
\newblock \doi{10.1121/1.2936367}

\bibitem{sabine_collected_1922}
W.C. Sabine, \emph{Collected Papers on Acoustics} ({Cambridge : Harvard
  University Press}, 1922)

\bibitem{nickolls_scalable_2008}
J.~Nickolls, I.~Buck, M.~Garland, K.~Skadron, in \emph{{{ACM SIGGRAPH}} 2008
  {{Classes}}} ({ACM}, {New York, NY, USA}, 2008), {{SIGGRAPH}} '08, pp.
  16:1--16:14.
\newblock \doi{10.1145/1401132.1401152}

\bibitem{myklebust_computing_2015}
T.G.J. Myklebust, Computing accurate {{Horner}} form approximations to special
  functions in finite precision arithmetic, arXiv:1508.03211 [cs, math]  (2015)

\bibitem{diaz-guerra_robust_2020}
D.~{Diaz-Guerra}, A.~Miguel, J.R. Beltran, Robust {{Sound Source Tracking Using
  SRP}}-{{PHAT}} and {{3D Convolutional Neural Networks}}, arXiv:2006.09006
  [cs, eess]  (2020)

\bibitem{luo_fasnet_2019}
Y.~Luo, C.~Han, N.~Mesgarani, E.~Ceolini, S.C. Liu, in \emph{2019 {{IEEE
  Automatic Speech Recognition}} and {{Understanding Workshop}} ({{ASRU}})}
  (2019), pp. 260--267.
\newblock \doi{10.1109/ASRU46091.2019.9003849}

\bibitem{luo_end--end_2020}
Y.~Luo, Z.~Chen, N.~Mesgarani, T.~Yoshioka, in \emph{{{ICASSP}} 2020 - 2020
  {{IEEE International Conference}} on {{Acoustics}}, {{Speech}} and {{Signal
  Processing}} ({{ICASSP}})} (2020), pp. 6394--6398.
\newblock \doi{10.1109/ICASSP40776.2020.9054177}

\bibitem{wang_neural_2020}
D.~Wang, Z.~Chen, T.~Yoshioka, Neural {{Speech Separation Using Spatially
  Distributed Microphones}}, arXiv:2004.13670 [cs, eess]  (2020)

\bibitem{mirbagheri_c-sl_2020}
M.~Mirbagheri, B.~Doosti, C-{{SL}}: {{Contrastive Sound Localization}} with
  {{Inertial}}-{{Acoustic Sensors}}, arXiv:2006.05071 [cs, eess]  (2020)

\bibitem{ceolini_evaluating_2020}
E.~Ceolini, I.~Kiselev, S.C. Liu, Evaluating {{Multi}}-{{Channel
  Multi}}-{{Device Speech Separation Algorithms}} in the {{Wild}}: {{A
  Hardware}}-{{Software Solution}}, IEEE/ACM Transactions on Audio, Speech, and
  Language Processing \textbf{28}, 1428 (2020).
\newblock \doi{10.1109/TASLP.2020.2989545}

\bibitem{ziegler_acoustic_2020}
J.D. Ziegler, H.~Paukert, A.K.a.A. Schilling, in \emph{Audio {{Engineering
  Society Convention}} 148} ({Audio Engineering Society}, 2020)

\bibitem{marvin182_room_2018}
Marvin182.
\newblock Room {{Impulse Response Generator}}.
\newblock https://github.com/Marvin182/rir-generator (2018)

\bibitem{hassani_multi-task_2017}
A.~Hassani, J.~{Plata-Chaves}, M.H. Bahari, M.~Moonen, A.~Bertrand,
  Multi-{{Task Wireless Sensor Network}} for {{Joint Distributed
  Node}}-{{Specific Signal Enhancement}}, {{LCMV Beamforming}} and {{DOA
  Estimation}}, IEEE Journal of Selected Topics in Signal Processing
  \textbf{11}(3), 518 (2017).
\newblock \doi{10.1109/JSTSP.2017.2676982}

\bibitem{markovich_multichannel_2009}
S.~Markovich, S.~Gannot, I.~Cohen, Multichannel {{Eigenspace Beamforming}} in a
  {{Reverberant Noisy Environment With Multiple Interfering Speech Signals}},
  IEEE Transactions on Audio, Speech, and Language Processing \textbf{17}(6),
  1071 (2009).
\newblock \doi{10.1109/TASL.2009.2016395}

\bibitem{qin_far-field_2019}
X.~Qin, D.~Cai, M.~Li, in \emph{Interspeech 2019} ({ISCA}, 2019), pp.
  4045--4049.
\newblock \doi{10.21437/Interspeech.2019-1542}

\bibitem{mosner_improving_2019}
L.~Mosner, M.~Wu, A.~Raju, S.H. Krishnan~Parthasarathi, K.~Kumatani,
  S.~Sundaram, R.~Maas, B.~Hoffmeister, in \emph{{{ICASSP}} 2019 - 2019 {{IEEE
  International Conference}} on {{Acoustics}}, {{Speech}} and {{Signal
  Processing}} ({{ICASSP}})} ({IEEE}, {Brighton, United Kingdom}, 2019), pp.
  6475--6479.
\newblock \doi{10.1109/ICASSP.2019.8683422}

\bibitem{severini_automatic_2019}
M.~Severini, D.~Ferretti, E.~Principi, S.~Squartini, Automatic {{Detection}} of
  {{Cry Sounds}} in {{Neonatal Intensive Care Units}} by {{Using Deep
  Learning}} and {{Acoustic Scene Simulation}}, IEEE Access \textbf{7}, 51982
  (2019).
\newblock \doi{10.1109/ACCESS.2019.2911427}

\bibitem{lehmann_matlab_2018}
E.A. Lehmann, Matlab {{Implementation}} of {{Fast Image}}-{{Source Model}} for
  {{Room Acoustics}}  (2018)

\bibitem{pavlidi_real-time_2012}
D.~Pavlidi, M.~Puigt, A.~Griffin, A.~Mouchtaris, in \emph{2012 {{IEEE
  International Conference}} on {{Acoustics}}, {{Speech}} and {{Signal
  Processing}} ({{ICASSP}})} (2012), pp. 2625--2628.
\newblock \doi{10.1109/ICASSP.2012.6288455}

\bibitem{alexandridis_capturing_2013}
A.~Alexandridis, A.~Griffin, A.~Mouchtaris, Capturing and {{Reproducing Spatial
  Audio Based}} on a {{Circular Microphone Array}}, Journal of Electrical and
  Computer Engineering \textbf{2013}, 1 (2013).
\newblock \doi{10.1155/2013/718574}

\end{thebibliography}

%\begin{thebibliography}{}
%%
%% and use \bibitem to create references. Consult the Instructions
%% for authors for reference list style.
%\end{thebibliography}

\end{document}